\documentclass[a4paper]{article}
\usepackage[utf8]{inputenc}           
\usepackage[USenglish]{babel}
\usepackage{graphicx}
\usepackage{subfigure}
\usepackage{url}

\begin{document}

\title{Numerical Calculations of Wake Fields and Impedances \hspace{10cm} of LHC Collimators'
  Real Structures
  \thanks{Work supported by HiLumi LHC Design
   Study, which is included in the High Luminosity LHC project and is partly
 funded by the European Commission within the Framework Programme 7 Capacities
Specific Programme, Grant Agreement 284404.}}

\author{O. Frasciello\thanks{oscar.frasciello@lnf.infn.it}, M. Zobov,
  INFN-LNF, Frascati, Rome, Italy}
\maketitle

\begin{abstract}
The LHC collimators have very complicated mechanical designs including movable jaws made of higly
resistive materials, ferrite materials, tiny RF contacts. Since the jaws are moved very close to the
circulating beams their contribution in the overall LHC coupling impedance is dominant, with respect
to other machine components. For these reasons accurate simulation of collimators' impedance becomes
very important and challenging.
Besides, several dedicated tests have been performed to verify correct simulations of
lossy dispersive
material properties, such as resistive wall and ferrites, benchmarking code results with analytical,
semi-analytical and other numerical codes outcomes.
Here we describe all the performed numerical tests and discuss the results of LHC collimators'
impedances and wake fields calculations.
\end{abstract}

\section{Introduction}

The Large Hadron  Collider (LHC) has a  very sophisticated collimation
system used to  protect the accelerator and  physics detectors against
unavoidable regular  and accident  beam losses \cite{assman2,assman1}.  The system  has a
complicated hierarchy composed of  the primary (TCP), secondary (TCS) and tertiary (TCT)
collimators and the injection protection collimators.

Since the  collimators are moved  very close to the  circulating beams
they  give the  dominant contribution  in the  collider beam  coupling
impedance,  both  broad-band  and  narrow  band.  The  electromagnetic
broad-band   impedance  is   responsible  of   several  single   bunch
instabilities  and  results in  the  betatron  tunes shift  with  beam
current, while the narrow band  impedance gives rise to the multibunch
instabilities and leads to vacuum chamber elements heating.

The  impedance related  problem  has been  recognized  already in  the
present LHC operating  conditions \cite{metrallhccoll} and is expected to  be even more
severe  for the  High  Luminosity LHC  upgrade \cite{hllhcdesrep}, where  one of  the
principal  key ingredients  for the  luminosity increase  is the  beam
current  increase.  For this  reason  the  correct simulation  of  the
collimator impedance becomes very important and challenging.

In order  to simulate the  collimators as  close as possible  to their
real  designs,  we used  CAD  drawings  including all  the  mechanical
details   as  inputs   for   the   high  performing,   parallelizable,
UNIX-platform FDTD  GdfidL code \cite{gdfidllink}.  A very fine mesh,  typically, of
several billions mesh  points, was required to reproduce  the long and
complicated structures, described in huge  .stl files, and to overcome
arising  numerical  problems.  In  order  to  be  sure that  the  code
reproduces   correctly  properties   of  lossy   dispersive  materials
(resistive walls,  ferrites) used in  the collimators we  have carried
our several dedicated numerical tests comparing the GdfidL simulations
with  available   analytical  formulae,  other  numerical   codes  and
semi-analytical mode matching techniques.

The only way to  afford such a huge computational task  was to use the
GdfidL dedicated  cluster at  CERN, engpara, which  has allowed  us to
study the wake fields and  impedances for several types of collimators
without using  any model  simplifications: secondary  collimators, new
collimators  with incorporated  beam position  monitors and  injection
protection collimators. In such circumstances, GdfidL wake fields computation
up to wake length of hundreds times the typical devices lengths ($\sim 1$m)
took several days or two weeks at maximum.

In this  paper we describe GdfidL tests of the resistive walls and ferrites
simulations, discuss the  calculated collimator  impedances comparing  the obtained
results with available experimental data.

\section{Resistive Wall Simulation Test}

\begin{figure}[!htb]
  \centering
  \includegraphics[scale=.4]{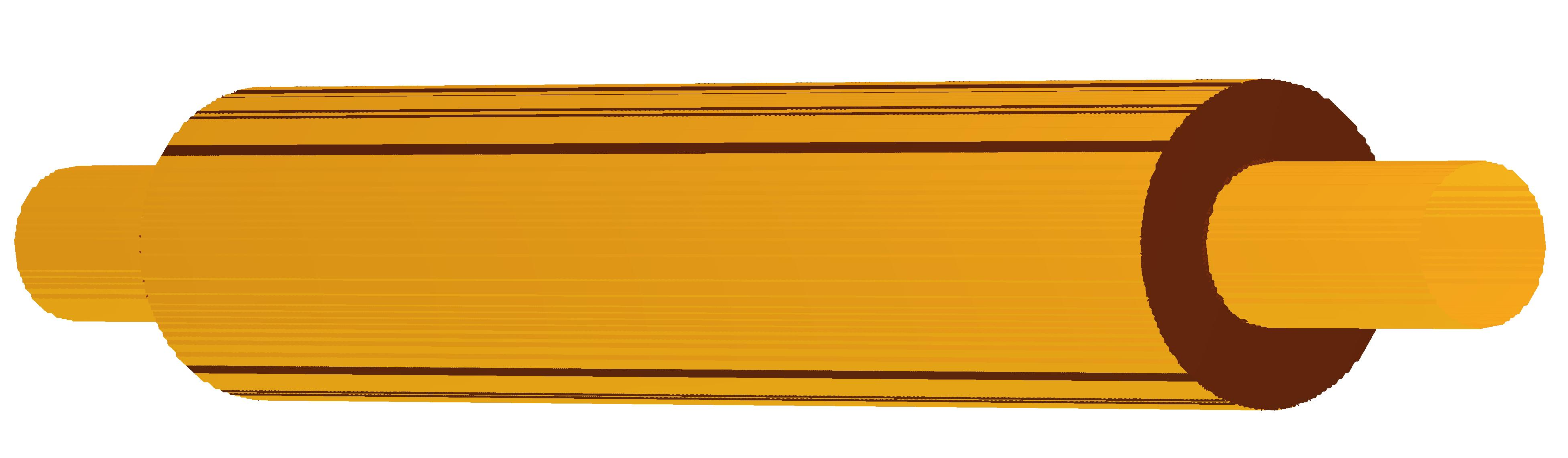}
  \caption{GdfidL model for the azimuthally symmetric beam pipe with resistive insert.
    The choosen length was
    $L=30$\,cm, the insert thickness $a=5$ mm, the pipe radius plus the insert thickness
    $b=10$ mm, and the electrical conductivity $\sigma_c$=$7.69\cdot 10^5$ S/m
    for Carbon Fiber Composite (CFC).}
  \label{fig:1}
\end{figure}

Only recently  a possibility to  carry out simulations  with resistive
walls (RW), implementing the impedance  boundary conditions, was made available
in GdfidL.  So
it  has  been  decided  to  perform a  numerical  test  comparing  the
simulation results  with known analytical formulas.   For this purpose
we calculated  both the longitudinal  and the transverse  loss factors
(the latter  known also as  kick factor)  of a Gaussian  bunch passing
inside  a  round  beam  pipe having  an  azimuthally  symmetric  thick
resistive insert.  The  insert was enough long in order  to be able to
neglect the  contribution of the insert  ends, as shown in Fig.~\ref{fig:1}.

\begin{figure}[!htb]
  \centering
  \includegraphics*[scale=0.5]{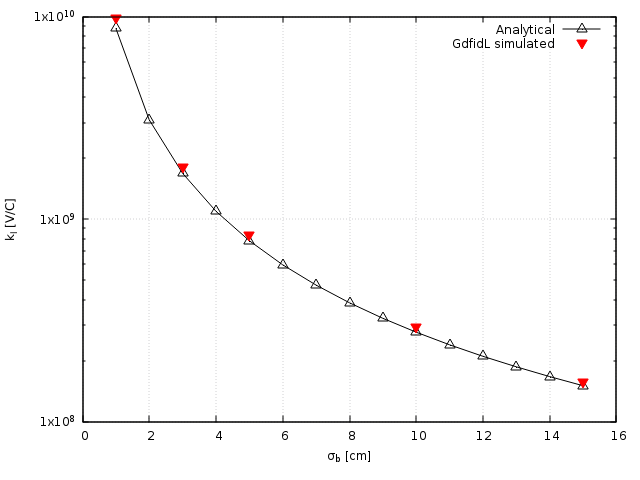}\label{fig:2a}\hfill
  \includegraphics*[scale=0.5]{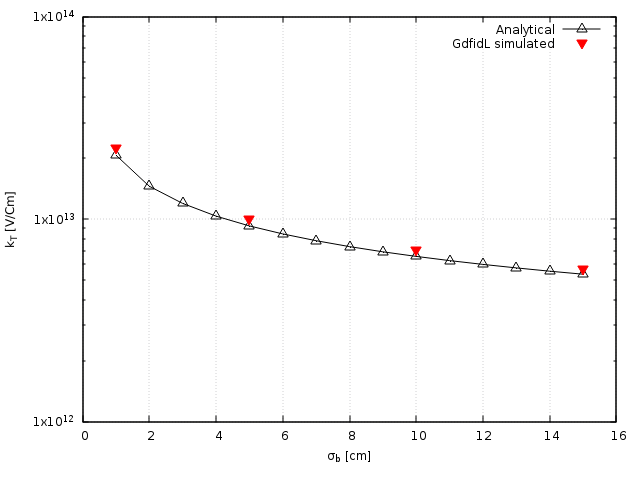}\label{fig:2b}
  \caption{Loss and kick factors benchmark between GdfidL and analytical formulas Eq.~(\ref{eq:1})
    and Eq.~(\ref{eq:2}).}
  \label{fig:2}
\end{figure}

In this case  the loss factors can be found analytically:
\begin{equation}
  k_{\parallel}=\frac{cL}{4\pi b\sigma_z^{3/2}}\sqrt{\frac{Z_0\rho}{2}}
  \Gamma\left(\frac{3}{4}\right),
  \label{eq:1}
\end{equation}
for the longitudinal one and
\begin{equation}
  k_{\perp}=\frac{cL}{\pi^2b^3}\sqrt{\frac{2Z_0\rho}{\sigma_z}}
  \Gamma\left(\frac{5}{4}\right)
  \label{eq:2}
\end{equation}
for the transverse one, where $c=2.997925\times 10^8$ m/s is the speed of light, $L$ is the length of
the pipe, $\rho=1/\sigma_c$ is the electrical resistivity, $\sigma_z$ the bunch length and
$\Gamma$ the Euler gamma function.
Figure~\ref{fig:2} shows a comparison  between the  analytical formulas  and the
numerical data. As it is seen the agreement is quite satisfactory.

\begin{figure}[!htb]
  \centering
  \includegraphics*[scale=0.45]{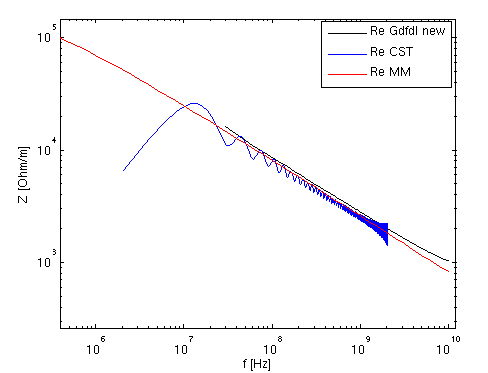}
  \caption{Dipolar transverse impedance benchmark between GdfidL, CST and MMM
  outcomes.}
  \label{fig:3}
\end{figure}

However,   the   loss   factors   are   somewhat   ``averaged''   values
characterizing the  beam impedance.  In order  to check  the impedance
frequency behavior the RW impedance  of the insert has been calculated
using the  semi-analytical  mode-matching method  (MMM) \cite{PhysRevSTAB.17.021001}.
In  turn,
numerically  the  impedance  till  rather  low  frequencies  has  been
obtained by  performing a Fourier  transform of  a long wake  behind a
long bunch obtained  by Gdfidl, and also by CST  for comparison. As it
is seen in Fig.~\ref{fig:3} also the impedance frequency behavior is reproduced
well by GdfidL.

\section{Ferrite Material Simulation Test}

In  order to  damp  parasitic higher  order modes  (HOMs)  in the  new
collimators with embedded  BPM pickup buttons, special  blocks made of
the TT2-111R lossy ferrite material are  used. For this reason we have
carried out  a comprehensive  numerical study to  test the  ability of
GdfidL  to  reproduce  frequency  dependent properties  of  the  lossy
ferrite  in calculations  of  wake fields,  impedances and  scattering
matrix parameters \cite{frasciello3}.

For  this purpose, we  have  a) simulated  a typical
 coaxial-probe measurement of
the  ferrite scattering  parameter $S_{11}$; b)  compared the  computation
results of  CST, GdfidL  and Mode  Matching Techniques by calculating
impedances of an azimuthally symmetric pill-box cavity filled with the
TT2-111R  ferrite  in  the  toroidal  region;  c)  benchmarked  GdfidL
simulations against  analytical Tsutsui model  for  a rectangular
kicker  with  ferrite   insert \cite{tsutsui1,tsutsui2} and  CST  simulations   for  the  same
device.

\begin{figure}[!htb]
  \centering
  \includegraphics[scale=0.8]{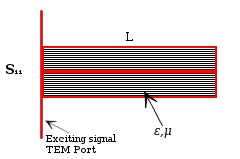}
  \caption{Coaxial probe measurement model for GdfidL $S_{11}$ simulations.}
  \label{fig:4}
\end{figure}

\begin{figure}[!htb]
  \centering
  \includegraphics[scale=0.4]{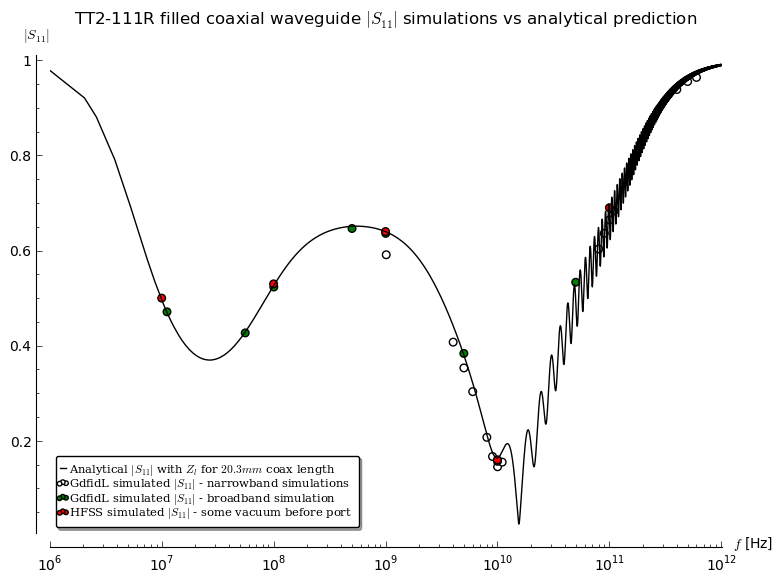}
  \caption{Reflection coefficient $S_{11}$ results for the arranged simulation setup.
  The solid line is the analytical trend from Eq.~(\ref{eq:s11}).}
  \label{fig:5}
\end{figure}

All the  comparative studies have  confirmed a good  agreement between
the  results obtained  by GdfidL  and  the results  provided by  other
numerical  codes, by  available analytical  formulas and  by the  mode
matching  semi-analytical approach.  As  an example,  Fig.~\ref{fig:4} shows  a
simplified  sketch of  a set-up  for the  ferrite material  properties
measurements: just a coaxial line filled with a ferrite material under
test. For  such a simple  structure the reflection coefficient $S_{11}$ is
easily measured and can be found analytically as in Eq.~(\ref{eq:s11}).
\begin{equation}
  S_{11}=\frac{\Delta\cdot \tanh(\gamma L)-1}{\Delta\cdot \tanh(\gamma L)+1},
\label{eq:s11}
\end{equation}
with $\gamma= j\omega\sqrt{\epsilon \mu}$ and $\Delta=\sqrt{\frac{\mu_r}{\epsilon_r}}$.
Figure~\ref{fig:5} shows the $S_{11}$ coefficient calculated for the TT2-111R material
in a very  wide frequency range, from  $10^6$ to $10^{12}$ Hz. As  it is seen,
despite the complicated $S_{11}$ frequency dependence the agreement between
GdfidL, HFSS and the analytical formula is remarkable.

\section{Impedance of LHC Run I TCS/TCT Collimators}

In the  2012 LHC  impedance model, collimators  played the  major role
($\sim 90\%$) over a wide frequency range, both for real and imaginary
parts,  but the  model was  essentially  based on  the resistive  wall
impedance of collimators, the resistive wall impedance of beam screens
and warm vacuum  pipe and a broad-band model  including pumping holes,
BPMs, bellows, vacuum valves and other beam instruments. The geometric
impedance  of collimators  was approximated  only by  that of  a round
circular taper \cite{Mounet:1451296}.

\begin{figure}[!htb]
  \centering
  \includegraphics[scale=.09]{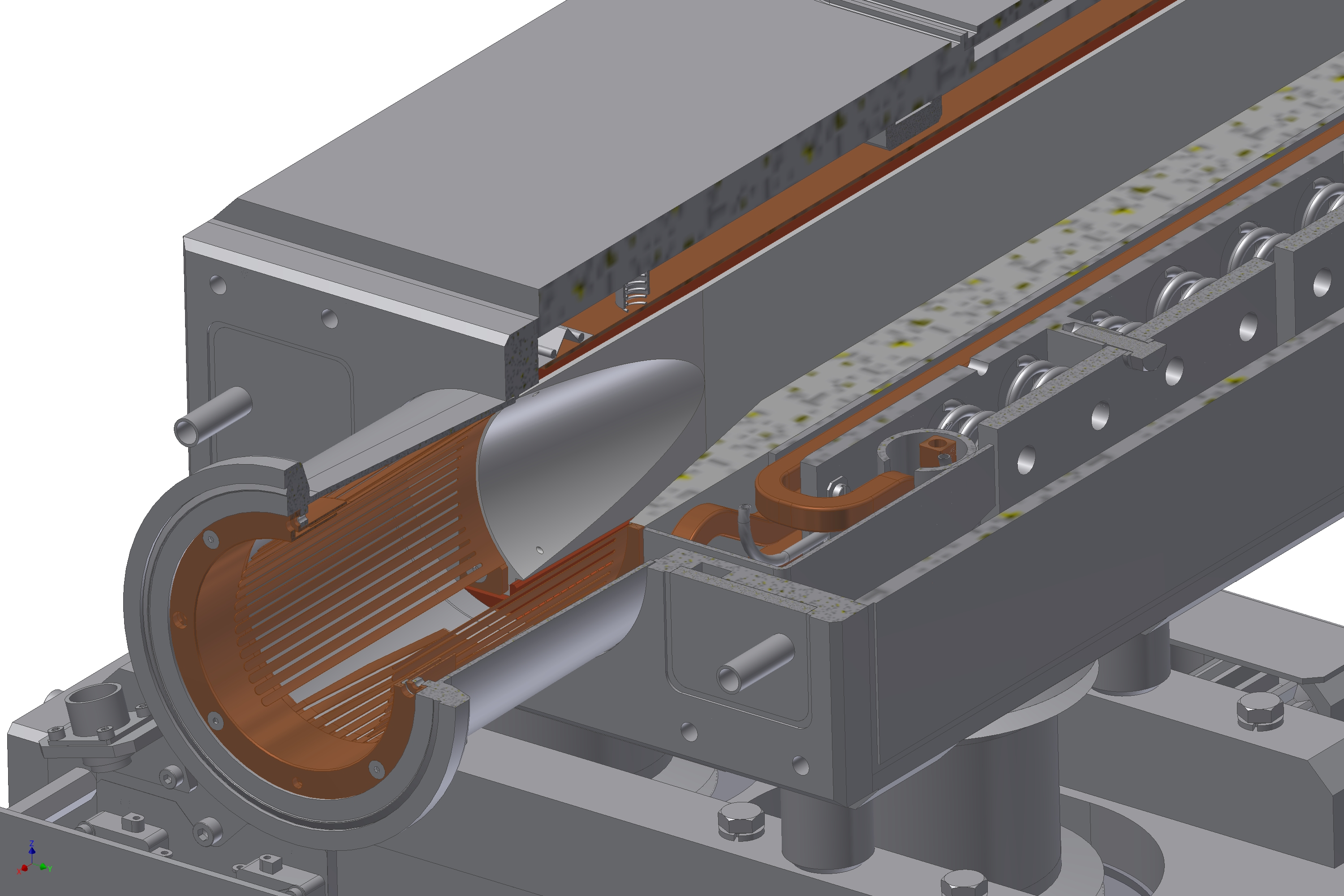}
  \caption{LHC Run I TCS/TCT collimator design.}
  \label{fig:10}
\end{figure}

However, several  measurements were performed in 2012 of the  total single
bunch tune shifts vs.  intensity, both  at injection and at 4 TeV, the
results coming out to  be higher than  predicted ones  with numerical
simulations by a factor  of $\sim 2$ at top energy and  of $\sim 3$ at
injection \cite{mounet2012beam}. This fact  led to the need  for an LHC
impedance  model refining  which,  first of  all,  required a  careful
collimator  geometric impedance  calculation.   For  this purpose,  we
carried out numerical  calculations of the geometric  impedance of the
LHC Run I TCS/TCT collimator, whose design is shown in Fig.~\ref{fig:10},
and  evaluated its contribution to the overall LHC impedance budget.

In order  to verify whether  the geometric collimator  impedance could
give  a  noticeable  contribution  to the  betatron  tune  shifts,  we
suggested to compare transverse kick factors due to the resistive wall
impedance  and the  geometric one,  showing that  the tune  shifts are
directly       proportional       to      the       kick       factors
\cite{frasciello1}.

\begin{figure}[!htb]
  \centering
  \includegraphics[scale=0.5]{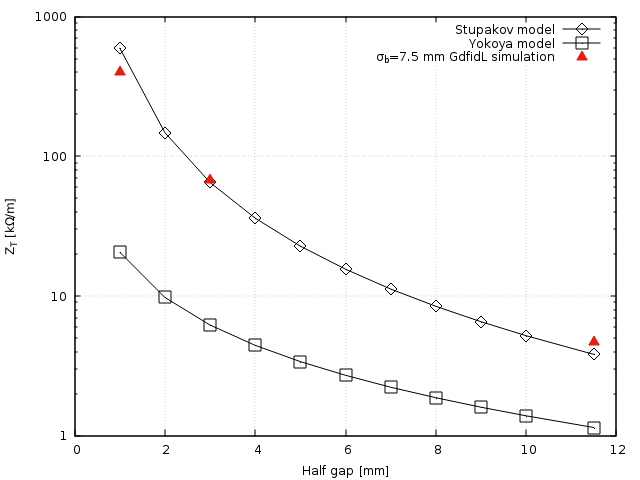}
  \caption{Effective transverse impedances of
  theoretical Stupakov flat taper model, Yokoya round taper model)
  and GdfidL simulations of TCS/TCT collimator, as a function
  of the jaws' half gap.}
  \label{fig:15}
\end{figure}

\begin{figure}[!htb]
  \centering
  \includegraphics[scale=0.5]{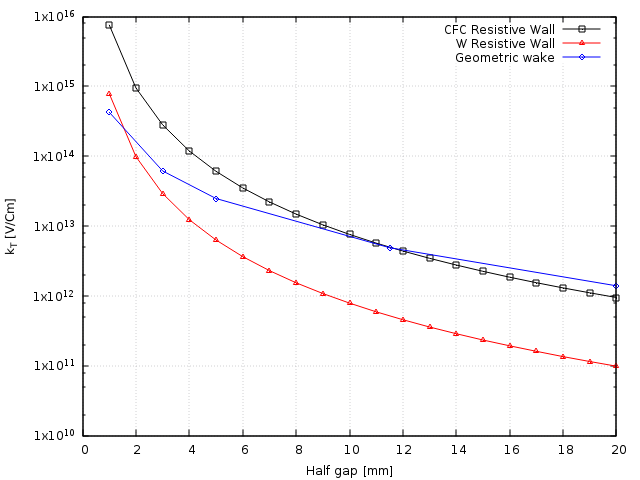}
  \caption{Comparison between geometrical kick factor and RW kick factors for CFC and W, as a
  function of the jaws' half gap.}
  \label{fig:16}
\end{figure}

 In Fig.~\ref{fig:15} and Fig.~\ref{fig:16} the main results about the
transverse broad-band impedance and kick factors are reported, showing
that the geometric impedance is better approximated by a flat taper model
than by a round taper one and that the  geometric contribution is not
negligible with respect to the  resistive wall one. In particular, for
CFC made collimator,  the geometrical kick starts to  be comparable to
resistive  wall one  at about  8  mm half  gap.  In turn,  for W  made
collimators,  the  geometrical  kick  dominates  almost  for  all  the
collimator gaps.

The study contributed to the refinement of the LHC impedance model. It
has also been shown that the geometrical collimator impedance accounts
for  approximately  $30\%$  of  the total  LHC  impedance  budget,  at
frequencies close to 1 GHz.

\section{Impedance of LHC Run II TCS/TCT Collimators}

\begin{figure}[!htb]
  \centering
  \includegraphics[scale=0.09]{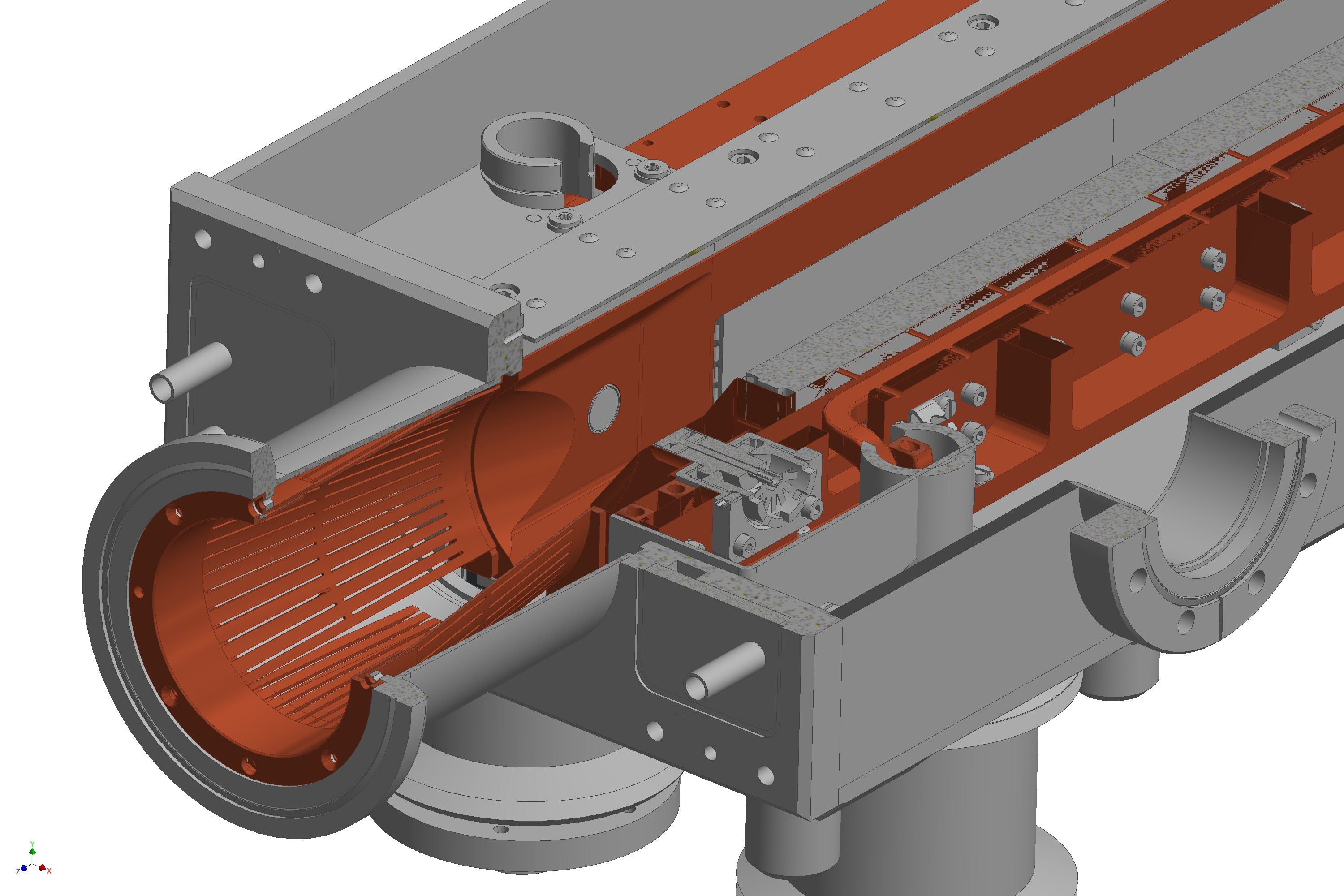}
  \caption{LHC Run II TCS/TCT collimator with embedded BPM pickup buttons.}
  \label{fig:17}
\end{figure}

During the last LHC Long Shutdown I (LSI), 2 TCS CFC and 16 TCT Tungsten (W) collimators
were replaced by new devices with embedded BPM pickup buttons, whose design is shown in
Fig.~\ref{fig:17}.
RF fingers were removed from the previous LHC Run I TCS/TCT design and HOMs damping
was entrusted to the TT2-111R ferrite blocks. By means of GdfidL broad band impedance simulations
of the new collimators' real structure, we gained the results for the kick factors in
Tab.~\ref{tab:1}, showing that an increase of about $20\%$ is expected
for the transverse effective impedance, with respect to  LHC RUN I type collimator's design.
\begin{table}[!htb]
  \centering
  \caption{Geometric Transverse Kick Factors Due to the Two TCS/TCT Geometries, Calculated at
    Different Half Gap Values}
  \newcommand{\VCm}{\ensuremath{\frac{\text{V}}{\text{Cm}}}}
  \begin{tabular}{ccc}
    \hline
    \hline
    & w/ BPM cavity        & w/o BPM cavity \\
    \hline
    Half gaps (mm)    & $k_T (\mathrm{\frac{V}{Cm}})$ & $k_T (\mathrm{\frac{V}{Cm}})$ \\
    \hline
    1                 & $3.921\cdot 10^{14}$    & $3.340\cdot 10^{14}$ \\
    3                 & $6.271\cdot 10^{13}$    & $5.322\cdot 10^{13}$ \\
    5                 & $2.457\cdot 10^{13}$    & $2.124\cdot 10^{13}$ \\
    \hline
    \hline
  \end{tabular}
  \label{tab:1}
\end{table}

\begin{figure}[!htb]
  \centering
  \includegraphics[scale=0.4]{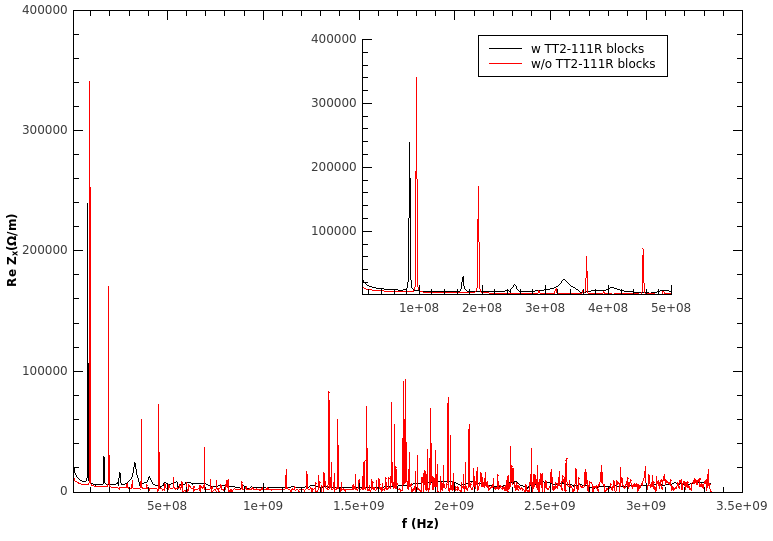}
  \caption{Real part of the impedance spectrum of LHC Run II TCS/TCT collimators, the inset layer
    focusing on low frequency HOMs.}
  \label{fig:20}
\end{figure}

In order to  study the impedance behaviour of the  new collimators and
the effect of the ferrite blocks on HOMs, we performed detailed GdfidL
wake fields simulations  of the whole real  structures.
We set into GdfidL input file the  finite conductivity  of W  and the  frequency-dependent
permeability  of  TT2-111R. As a first result,  an  overall
impedance damping feature was shown to be proper of the structure with resistive W
jaws plus ferrite blocks at  all frequencies \cite{frasciello4}, as clearly visible from the plot
in Fig.~\ref{fig:20}. There, the red curve represents the collimator simulated as a whole Perfect
Electrical Conductor (PEC), without any resistive and dispersive material, while the black one
represents the real collimator with W jaws and ferrite blocks. The effect of ferrite results also in
the shift of HOMs characteristic frequencies toward lower frequencies. As an example, the first
HOM frequency shifts from $\approx 95$ MHz to $\approx 84.5$ MHz, at exactly the same frequency
measured experimentally at CERN with loop technique \cite{biancacci1}.
It is clear that the computed  impedance
spectrum  resolved  very   well   the  low   frequency  HOMs,   whose
characteristic frequencies are in excellent agreement with those found experimentally.
Moreover, under these simulation circumstances, the computed shunt resistance of the first HOM at
$\approx 84$ MHz is in remarkably agreement, within a factor of $2$, with that measured
experimentally with the wire technique at CERN \cite{biancacci2}, being
$R_s^{sim}\approx 237$\,k$\Omega/$m and $R_s^{meas}\approx 152$\,k$\Omega/$m \cite{frasciello5}.

\section{Conclusions}

Calculations of wake fields and beam coupling impedance have been performed for the LHC
TCS/TCT collimators, by means of GdfidL electromagnetic code. We performed, for the first time
in the field of impedance
computations, a complete and detailed simulation campaign of collimators' real structures,
including the properties of real and lossy dissipative materials.

For LHC Run I collimators, the comparison of the transverse kick factors calculated for
five different jaws' half gaps, has shown that the geometric
impedance contribution is not negligible with respect to the resistive wall one.
The study has contributed to the refinement of the LHC impedance model, as a result of
 the geometrical collimator impedance accounting for approximately $30\%$ of
the total LHC impedance budget, at frequencies close to 1 GHz.

The performed  numerical tests  have confirmed that  GdfidL reproduces
very  well the  properties  of the  lossy  dispersive materials.   The
simulation test results for the resistive walls and the lossy ferrites
are in  a good  agreement with available  analytical formulae  and the
results  of other  numerical  codes and  semi-analytical models.
 The
tests have made  us confident in the results of  our impedance studies
carried  out  for the  real  structures  of  the  new Run  II  TCS/TCT
collimators  with incorporated  BPMs. Several  important results  have
been obtained conducting these studies. First, we found  that there are
no dangerous longitudinal higher order  modes till about 1.2 GHz. This
is  important for  the heating  reduction  of the  collimators in  the
multibunch regime (for the nominal  LHC bunches 7.5 long). Second, the
TT2-111R  ferrite  resulted  to  be very  effective  in  damping  both
longitudinal and transverse parasitic  modes for frequencies above
500  MHz. However,  the modes  at lower  frequencies are  less damped,
residual
transverse  HOMs  at frequencies  around  100  MHz  and 200  MHz  with
non-negligible  shunt  impedances   still  existing.   The  calculated
frequencies of  the modes  are in remarkable  agreement with  the loop
measurements.  The shunt impedances  of the modes obtained numerically
agrees within  a factor of  2 with the  experimental data of  the wire
measurements  performed at  CERN. Finally,  the broad-band  transverse
impedance of the new LHC Run II double taper collimators are evaluated
to be approximately by 20\% higher with respect to that of the LHC Run
I TCS/TCT collimators.

\section{Acknowledgments}
We are grateful to W. Bruns for his invaluable support.

We would like to thank also
the CERN EN-MME and BE-ABP departments, for the providing of the collimators' CAD designs and
S. Tomassini, of INFN-LNF, for his accurate handling and adjusting the CAD designs to serve as
inputs for GdfidL simulations.
In particular special thanks are addressed to the E. Metral, on behalf of the whole CERN LHC
impedance group, for the support and
profitable discussions, to N. Biancacci for the MMM simulations' data and, together with
F. Caspers, J. Kuczerowski, A. Mostacci and B.Salvant for the information on
the collimators' impedance measurements side.

\bibliography{refarxiv}
\bibliographystyle{unsrt}
\null
\end{document}